\input{aipcheck}

\documentclass[
    final            
  ]
  {aipproc}

\usepackage{natbib}
\layoutstyle{6x9}

\begin{document}

\title{X-ray Signatures of Circumnuclear Gas in AGN}

\classification{98.54}
\keywords      {galaxies:Seyfert, X-rays}

\author{T.J.Turner}{
  address={Dept. of Physics, University of Maryland Baltimore County, Baltimore, MD 21250}
}

\author{L.Miller}{
  address={Dept. of Physics, Oxford University, Denys Wilkinson Building, Keble Rd.,Oxford OX1 3RH, U.K.}
}

\author{M.Tatum}{  
address={Dept. of Physics, University of Maryland Baltimore County, Baltimore, MD 21250}
}

\begin{abstract}

X-ray spectra of AGN are complex. X-ray absorption and emission
   features trace gas covering a wide range of column densities and
   ionization states.  High
   resolution spectra show the absorbing gas to be outflowing,
   perhaps in the form of an accretion disk wind. The absorbing complex shapes the form of the
   X-ray spectrum while X-ray reverberation and absorption changes explain the  
   spectral and timing behaviour of AGN. We discuss recent
   progress, highlighting some new results and reviewing the
   implications that can be drawn from the data.

\end{abstract}

\maketitle


\section{Introduction}

Spectroscopic data from the X-ray, UV and optical regimes have 
 established the presence of significant columns of material 
 outflowing from supermassive black holes  in 
  active galactic nuclei (AGN).  Studying these outflows offers insight
 into the observed relationship between the black hole and its host galaxy 
\citep[e.g.][]{gebhardt00a}.  Given the short duty-cycle of accretion that 
seems to exist for AGN \citep{king10a},  epochs of significant nuclear outflow must occur for  AGN to 
build the supermassive black holes observed in the Universe.  
The most promising models for X-ray outflows consider mass loss via an
 accretion disk wind \citep[e.g.][]{sim10a}.  The X-ray band carries 
 information about the material flow very close to the black
 hole, and offers important constraints on fundamental parameters of
 the system, such as the wind launch radius and acceleration
 mechanism. 

 The spatial resolution of current X-ray instruments does
 not allow us to image the innermost regions of AGN directly, 
However, recent long X-ray observations,
 particularly those afforded by {\it Suzaku} and {\it XMM-Newton} have
 allowed a significant step forward using spectroscopic and timing analyses. 
 Here, we review our current understanding of AGN absorber systems, 
highlighting key contributions from the {\it Suzaku}
satellite.

\section{X-ray Absorption in AGN}

 X-ray grating spectroscopy has allowed more than an order
of magnitude improvement in spectral resolution at 1 keV, and a factor
$\sim 4$ improvement at 6 keV, compared with CCD
instruments.   Accurate measurement of the energies of soft-band absorption lines such as  those from 
O, Ne, Mg, Si and S, first traced relatively low ionization gas, outflowing at hundreds to thousands of
km\,s$^{-1}$ \citep{kaspi02a, blustin05a}. Such so-called 'warm absorbers' are common in AGN. 
The best-studied, local AGN reveal multiple zones of 
absorbing gas, whose signatures can be separated with grating data. 
Application of photo-ionization models can yield the 
ionization-parameter for the gas, and an estimate of the product of
density and radial distance from the source. In principle, the gas density may be
estimated using sensitive triplet line ratios, however, in practice
it has proved  difficult to obtain interesting constraints. Absorber variability can also be
used to constrain the gas density via consideration of the recombination
timescale, although attempts to date have been limited by scant variability measurements. 

 Estimates of the gas location  vary. Some new work considers reverberation analysis 
within the X-ray band, allowing a constraint on the location of 
reprocessing gas out of the line of sight.  A few cases studied to date place the reprocessor 
tens to hundreds of gravitational radii from the nuclear source \citep{miller10a}. 
Other methods include measurement of the gas ionization parameter that, combined with 
 velocity constraints and  
consideration of reasonable limits on the gas geometry, yield radial estimates 
that range from within the optical broad-line-region (BLR)  to 
outside the putative molecular torus \citep{blustin05a}. 

An important development has been the discovery of deep K-shell
 absorption lines from  H-like and He-like species of
 Fe \citep[e.g.][]{kaspi02a,reeves04a,turner08a}. {\it Suzaku} has been effective at detection of these lines, 
owing to its high sensitivity at 6 keV. Importantly, 
{\it Suzaku} data yielded detections of Fe\,{\sc xxv} and Fe\,{\sc xxvi} in MCG--6-30-15 (Fig.~1),  
with outflow velocity  $\sim 1800$\,km\,s$^{-1}$ \citep{young05a,miller08a}. 
In general, measured line
depths indicate the presence of very high columns of gas, 
$N_{\rm H} \sim 10^{23} - 10^{24}$\,atoms\,cm$^{-2}$  
{\it in the line-of-sight}, with  velocities typically much higher than those 
detected in the soft X-ray 
regime \citep[e.g.][]{risaliti05a,miller07a,turner08a}. 
A sample study has found velocities up to 
$\sim 0.3$~c for these high-ionization absorbers \citep{tombesi10a}  
with some systems transporting significant mechanical energy from the 
nucleus to the  host galaxy \citep{pounds09a}.

\begin{figure}
  \includegraphics[height=.25\textheight,angle=-90]{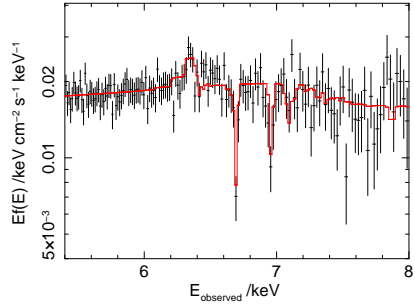}
\caption{Unfolded Chandra HEG spectral data of MCG--6-30-15 from 2004.
The spectral model is shown as a solid line. 
See text and \citet{miller08a} for details.}

\end{figure}

Systems detected to date span six orders of
magnitude in ionization parameter, and five orders of magnitude in
integrated column density. In summary, we have detected absorbing
gas over the entire range of parameter-space to which current X-ray 
instruments are sensitive.

\section{Spectral Variability}

In light of the obvious importance of complex absorption in 
 MCG--6-30-15, \citet{miller08a} compiled
data from  
 {\it Chandra}, {\it XMM-Newton}  and {\it Suzaku} to conduct
  the most comprehensive broad-band analysis (0.5-45 keV) of a single AGN 
to date. Spectral
variations were decomposed using the SVD-PCA method, showing that  spectral variability in 
 MCG--6-30-15 could be fully described with a simple variations of one partial-covering zone of the 
complex X-ray absorber. 
The  variable-covering absorber suggested for MCG--6-30-15 is of general interest, 
as it accounts for the spectral variations seen in a number of 
AGN \citep{miller07a,miller08a}. 

 In the case of a few sources, such
as NGC 3516 \citep{turner08a}, the data require an additional
component of variability in the  high-state that  may either
represent a clear view of the continuum variability, or else may provide
evidence for dense clumps of gas ($N_{\rm H} > 10^{25}$\,cm$^{-2}$) in the 
absorber system  \citep{turner11a}. Further observations using  {\it Suzaku} are key to breaking this ambiguity, 
providing the only method of tracking AGN behavior simultaneously over 0.5 - 50 keV and the only direct 
way to probe the importance of Compton-thick gas in the absorber complex.  

\section{Compton-thick gas in the line-of-sight}

{\it Suzaku} observations of 1H 0419--577 and PDS~456 have been key in
reshaping our view of AGN.  For both sources, a marked `hard excess' of counts was
detected in the PIN data relative to the predicted flux based on model
fits below 10\,keV \citep{turner09b,reeves09a}. The  high PIN-band flux
(Fig.~2) can be modeled with 
a layer of Compton-thick  gas 
(taken as $N_H=1.25 \times 10^{24} {\rm cm^{-2}}$) that covers $\sim 90\%$
of the continuum (after correction for scattering losses).  {\bf What is
remarkable is that this Compton-thick gas exists in the
line-of-sight to the nucleus}. The type 1 classification of 1H 0419--577 
implies that the Compton-thick gas lies within the BLR, or that
the BLR is observed through holes in the absorbing medium.  Both  
possibilities challenge current models for AGN and necessitate a reconsideration of the 
true intrinsic luminosity distribution of the population.

\begin{figure}
  \includegraphics[height=.3\textheight,angle=0]{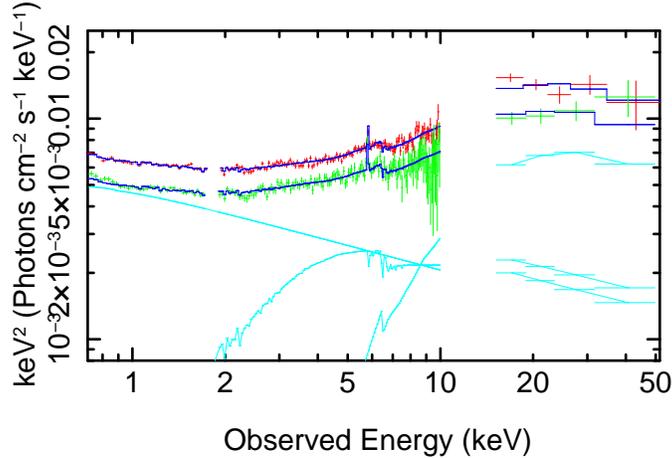}
  \caption{{\it Suzaku} data for 1H0419--577 from 2007(red) and 2010 (green). 
The solid dark blue line
shows the total model fit for each epoch. Pale blue lines show 
the model components, representing three different
sight-lines through a complex absorber.  Aside from the normalization, the 
 spectral model is the
same for 2007 and 2010 data, and so only component lines from 2010 are shown, for
clarity. Note a slight offset of the model lines in the PIN band,
owing to an 18\% re-normalization factor that must be applied using the most recent 
calibration.} 
\end{figure}

These observations prompted a follow-up study.   The {\it Swift} BAT survey provides a survey of the X-ray
sky that is unbiased up to column densities $N_{\rm H} \sim 10^{24}{\rm
cm^{-2}}$.   Tatum et al.
(in prep) selected a sample of type 1 AGN from the 58-month BAT 
catalog. To understand the sample, we required simultaneous medium and hard-band X-ray data, and so the BAT 
 list was cross-correlated with the {\it Suzaku} archive, yielding a sample comprising 65 observations 
of 50 objects. The energy flux was measured in the 15--50\,keV and 2--10\,keV 
bands using the simultaneous XIS and PIN data.  The results are shown in Fig.~3. 
The weighted mean ratio is shown as a black solid
line; surprisingly, 1H0419--577 is consistent with the sample mean,  
having a hardness ratio $\sim 1.4$.  The red line represents 
the sum of a powerlaw  ($\Gamma=2.1$) plus reflection from a thin disk of neutral
material subtending  $2\pi$ steradians to the illuminating continuum. To
characterize the reflector we used the {\sc pexrav} model in {\sc xspec}, 
assuming no cut-off of the continuum, 
an inclination  angle of 60$^{\rm o}$  and solar abundances.
 The distribution of measured  ratios are 
generally much harder than that  model.

The solid orange line represents the 
hardness ratio for the same reflector, this time assuming the illuminating 
continuum is hidden from view (pure reflection). While a pure reflection spectrum might be expected for type 2 AGN, such an explanation is at odds with a sample of type 1 AGN given the current model for these systems. 
Markers on the 
right side of the plot represent the ratios expected for the case 
where a neutral column of Compton-thick gas 
partially covers the continuum source. Several 
covering fractions are denoted: 98\% (green), 90\% (dark blue), 70\%
(light blue) and 50\% (magenta).  In the context of this simple
partial-covering model, a remarkable 70\% of the sample are found to
have significant covering by Compton-thick gas in the
line-of-sight. While some Compton-thick AGN have been identified in
previous BAT-selected samples of AGN \citep{winter09a}, our approach, using only the
simultaneous constraints provided by the {\it Suzaku} XIS and PIN, has
shown the enormous importance of partially-covering Compton-thick gas in the
local population of type 1 AGN.

\begin{figure}
  \includegraphics[height=80mm,width=100mm,angle=0]{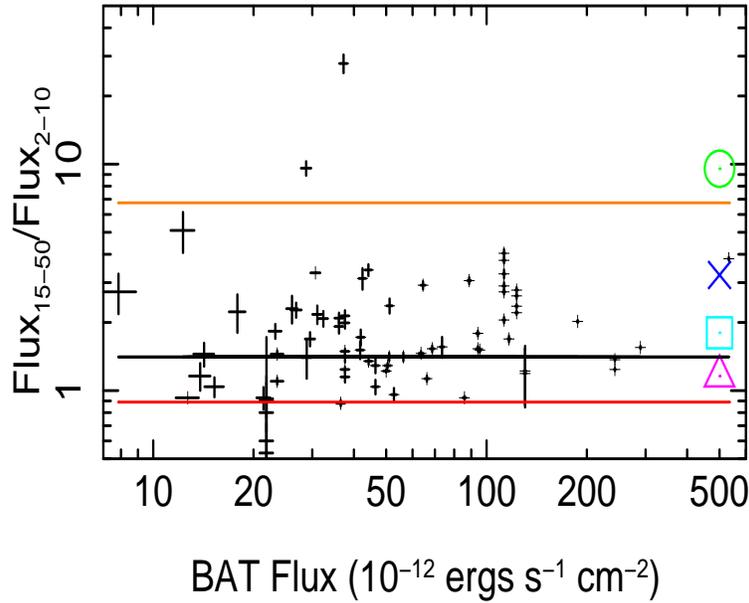}
  \caption{ {\it Suzaku} hardness ratio plotted against BAT flux, using the simultaneous energy density
fluxes measured from {\it Suzaku} PIN/XIS. The sample comprises type 1 AGN selected from
the 58-month BAT catalog cross-correlated with the {\it Suzaku} 
archive (Tatum et al in prep). The black line represents the weighted mean hardness for the sample; the red 
line represents the hardness ratio expected from a power-law  
reflected from a standard thin disk subtending 2$\pi$ steradians on the sky; the orange line is the ratio expected from pure, neutral reflection. The green, dark blue, light blue and magenta markers represent the hardness ratios expected from a partially-covered power-law, with neutral gas and covering fractions 98\%, 90\%, 70\% and 50\%. The continuum power-law is  taken to be $\Gamma=2.1$, for all models. 
}
\end{figure}

 It is clear that studies below 10 keV have greatly underestimated the 
intrinsic luminosity of a large fraction of the local AGN population (Tatum et al in prep.) and that 
a reconsideration of the structure of the inner regions of AGN is unavoidable. 
 {\bf Results such as this illustrate the enormous value 
the broad-band coverage of {\it Suzaku} offers to the astronomical community.}

\section{Broad Components of Fe K$\alpha$ emission}

Compelling evidence for Compton-thick gas in the line-of-sight, combined with 
the certainty that the X-ray absorber is outflowing, has
motivated detailed calculation of the  signatures 
from a Compton-thick flow \citep{sim08a,sim10a}. Consideration of high wind densities 
predicts a strong Fe K emission line should be observed from the wind. This line is broadened  by a
 combination of electron scattering and velocity effects. 
Calculations assume the wind is launched beyond $\sim 30\, r_g$ and so relativistic blurring does not 
contribute to the predicted line width, indeed the predicted line widths are not extreme, having typical values  
FWHM$\sim 1$\,keV. Case studies of Mrk~766 \citep{sim08a} and PG1211$+143$ \citep{sim10a} showed
that a Compton-thick  wind can describe  the 
Fe\,{\sc xxv} and Fe\,{\sc xxvi} absorption lines and the broad Fe
emission component with a single flow.

\begin{figure}
  \includegraphics[height=.22\textheight,angle=0]{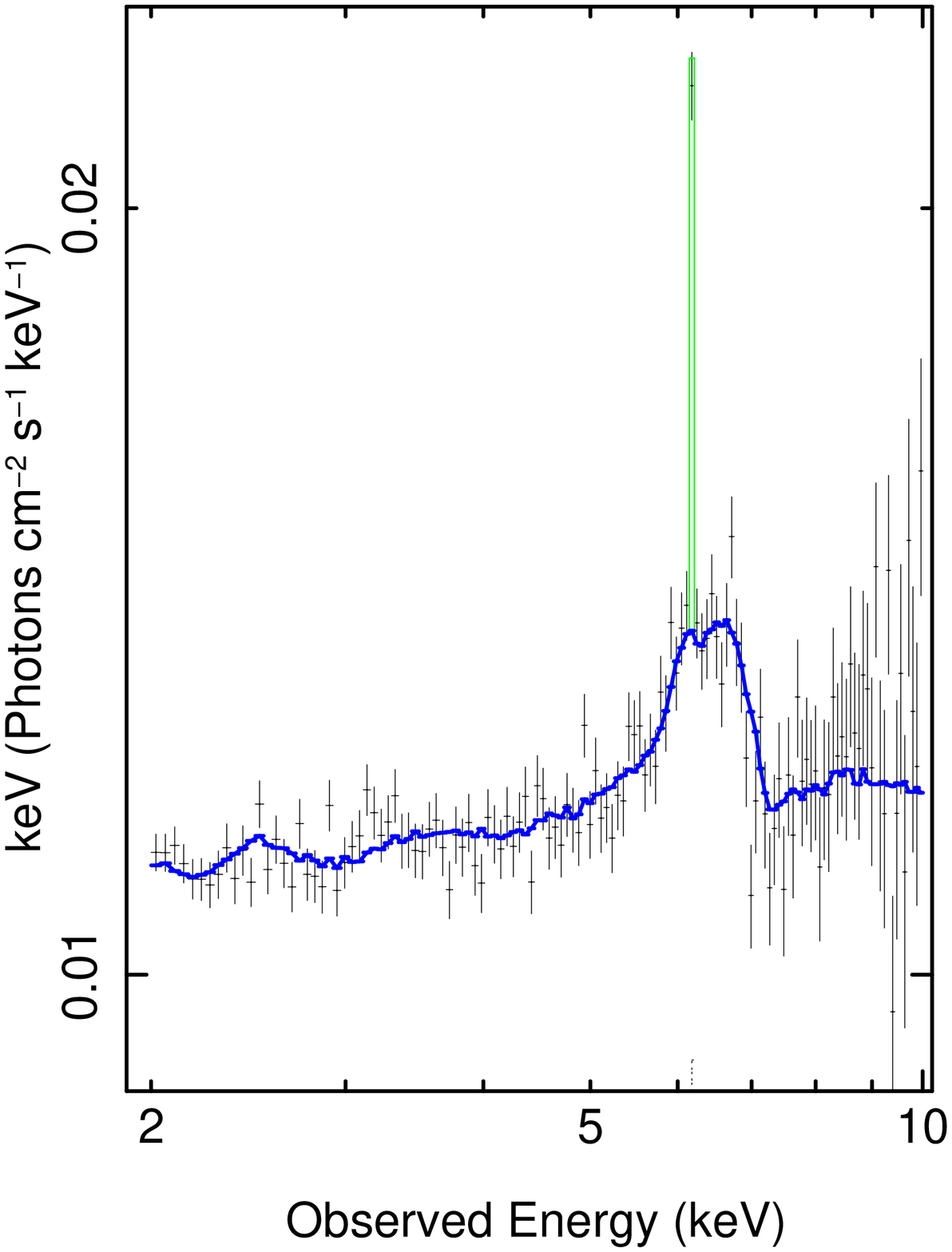}
  \includegraphics[height=.22\textheight,angle=0]{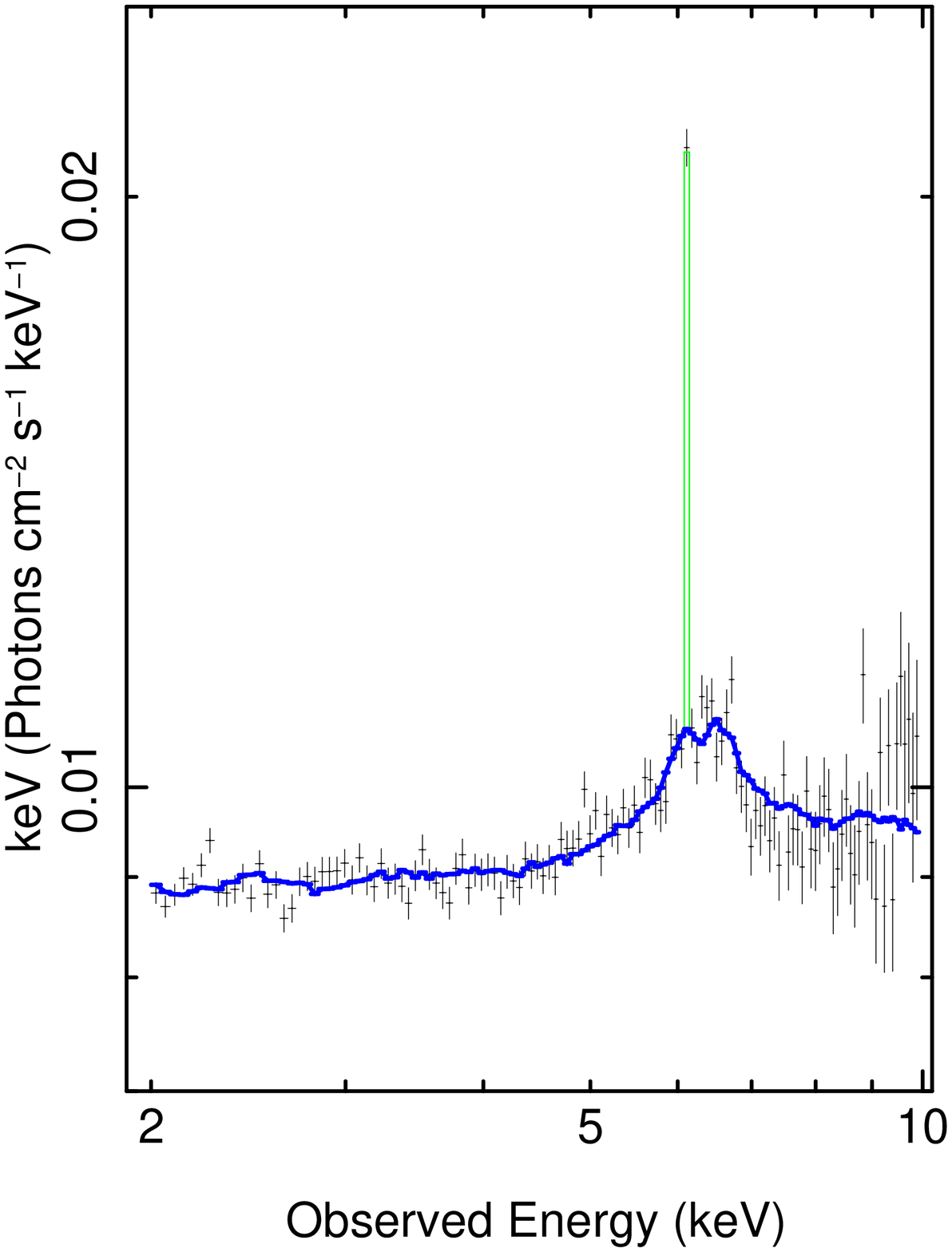}
  \includegraphics[height=.22\textheight,angle=0]{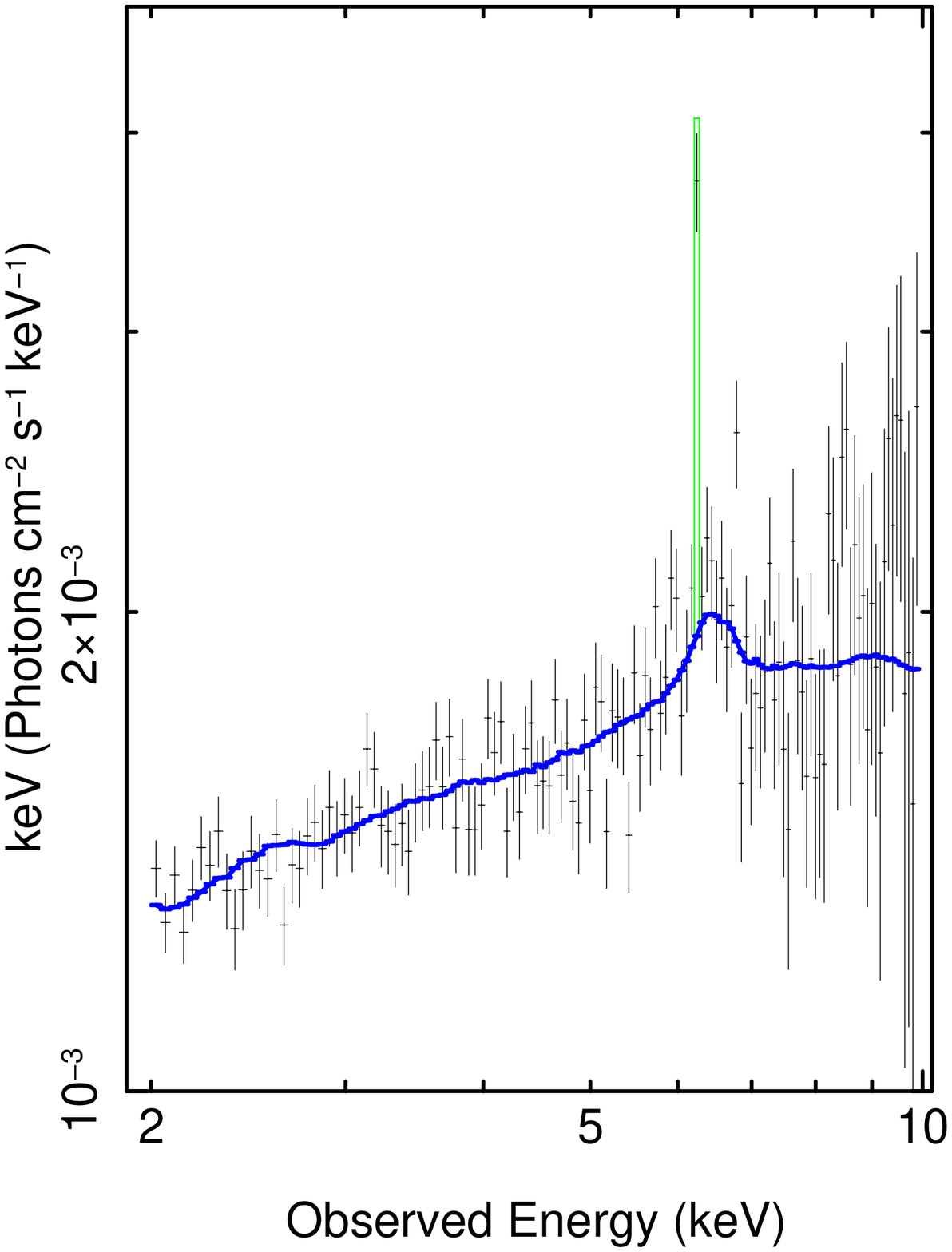}
  \includegraphics[height=.22\textheight,angle=0]{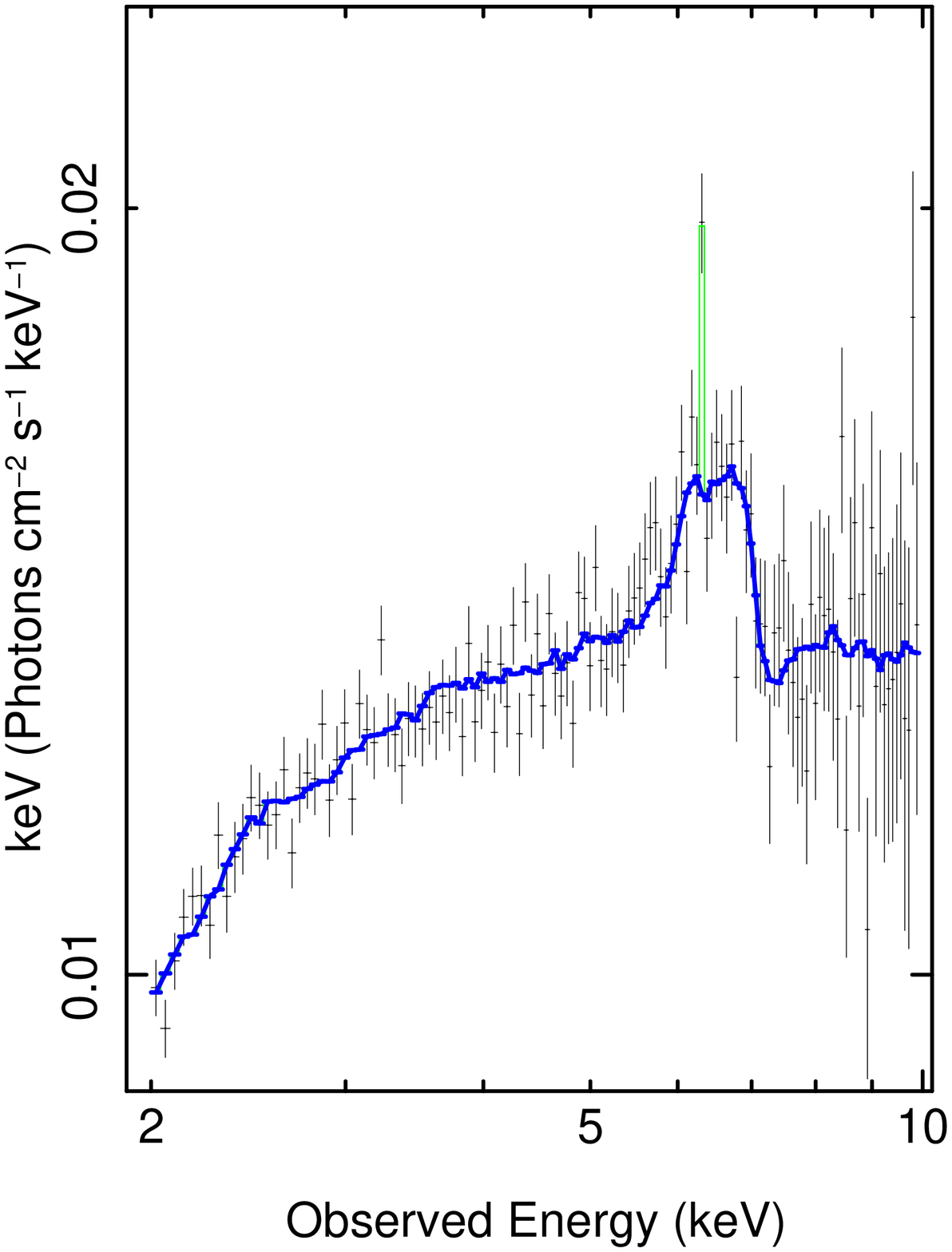}

  \caption{{\it Suzaku} XIS data for a sample of unobscured AGN. Left to right: Ark 120, Fairall-9, MCG--02-14-009 and Swift$2127.4+5654$. The solid blue line shows the 
 Compton-thick wind component of the model, the green line is a narrow emission component (see text for details).
}
\end{figure}

To follow up on those results,  Tatum et al (in prep.) have taken an exploratory sample of local
AGN \citep[see][]{patrick11a} that possess 
moderately-broad components of Fe K emission. 
{\it Suzaku} XIS data were fit over 2--10\,keV using a powerlaw continuum modified by a 
disk wind, plus Galactic absorption. Fig.~4 shows some examples of the results obtained. 
The solid blue 
line represents the continuum power-law convolved with absorption
and scattering effects from the wind (convolved with the effect of Galactic absorption). 
The green line shows 
the narrow component of Fe\,K$\alpha$ emission that was required in all
sources, likely representing reprocessing in distant material, not
necessarily associated with the wind. 
The wind model provided a satisfactory fit to all of the broad lines observed, indicating that winds may explain 
both the dominant absorption and emission properties of Seyfert type AGN (Tatum et al in prep.).

\section{Timing Analysis}

Measurement of reverberation signals is a powerful technique for
constraining the size-scale of astronomical systems.  
Reverberation signatures are measured by cross-correlating two time series and searching 
for time delays between them. To perform this technique within the X-ray band, one must 
understand the contributions of direct and reprocessed signatures in each of the bands considered. 

\citet{miller10a,miller10b} have 
developed a maximum-likelihood approach to investigate time delay signatures in Fourier space.
This method  finds 
the power-spectra and cross-power-spectra that best fit the time-domain data, with 
full consideration of data sampling and noise, and using estimated measurement uncertainties
that account for the variance of the source from observation to observation. 

A key part of the timing analysis consists of construction of a ``lag spectrum'' that  shows 
the time lags between two time series as a function of the frequency of the source variation. 
Analysis of the lag spectrum yields information about the geometry and size scale of the reprocessor. 
Consider  a thin spherical shell of material surrounding an active nucleus, where the 
directly-viewed continuum  is emitted from a central point source and measured in
one time series, while light scattered from the shell is measured in a second time
series.  If the shell has holes  ``partial-covering'' in the sight-line to the observer, and also on the 
opposite side of the continuum source from the observer, then the gas appears somewhat like a ring seen end-on to the observer.  
One can construct the lag spectrum for such a reprocessor. At high frequencies, where the variability mode  
is significantly shorter than the light travel time across the shell, scattering from different 
parts of the shell add incoherently over a wide range of phases, so the observed net time
delay tends to zero. One also observes some oscillations that depend on the details of the shell structure.  
At low frequencies, where the source variability period is much longer than the light travel time across
the shell, the scattered light adds mostly coherently and  a mean time delay is observed 
for the shell.  At a frequency where the time period equals the light travel time
across the shell, a sharp transition is observed  between these two regimes and the measured 
time delay may become  negative for some ranges of frequency. While all the
scattered light signals actually are delayed,  the negative lag appears  in Fourier space if the phases wrap around 
 by more than $2\pi$ radians.   Thus, the detailed oscillatory structure in the lag spectrum 
tells us about the  structure, geometry and size-scale of the reprocessor. Note that if the shell were symmetric and full-covering, the 
time delays oscillate but stay positive.

\begin{figure}
  \includegraphics[height=.22\textheight,angle=-90]{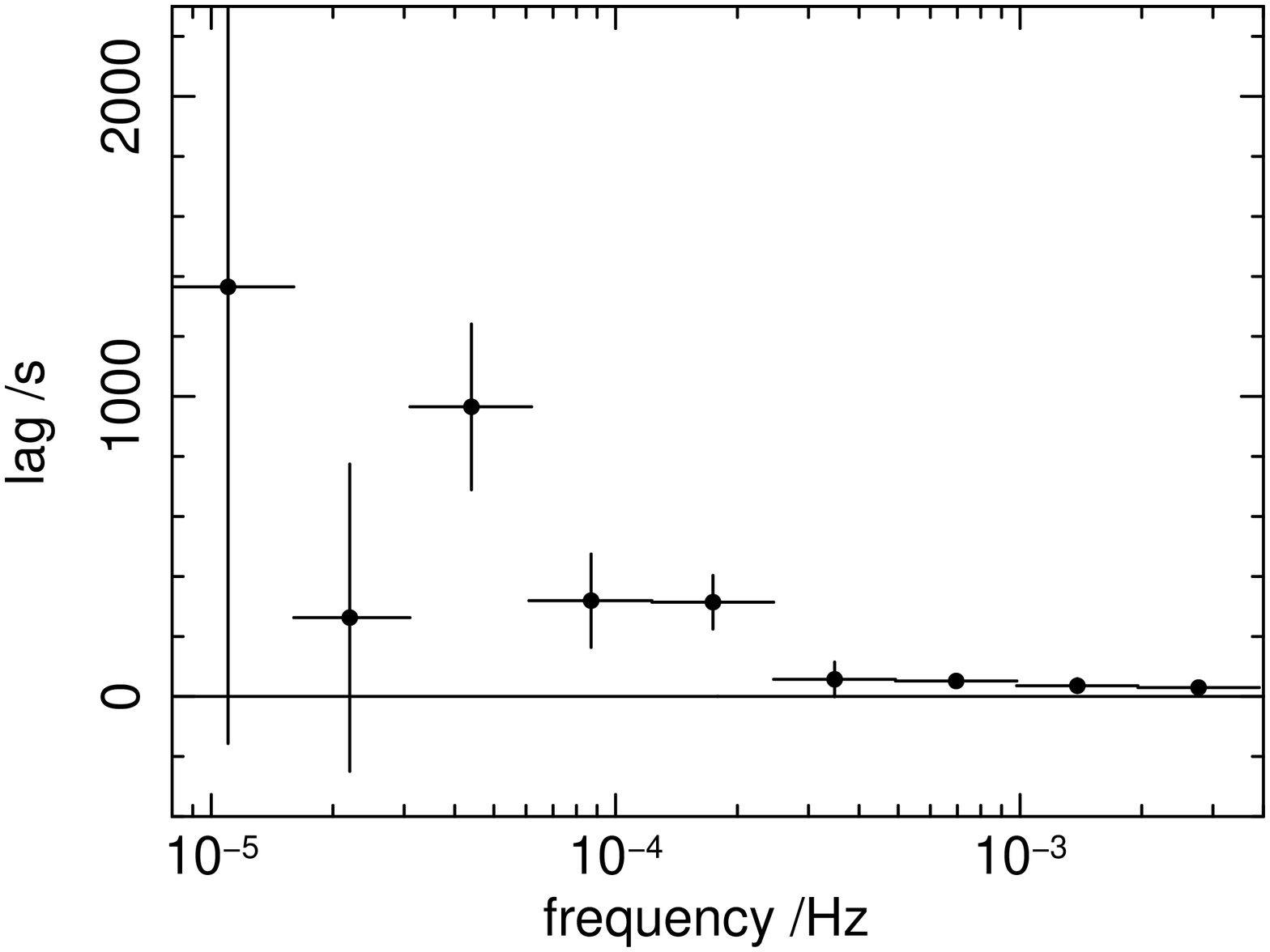}
  \includegraphics[height=.22\textheight,angle=-90]{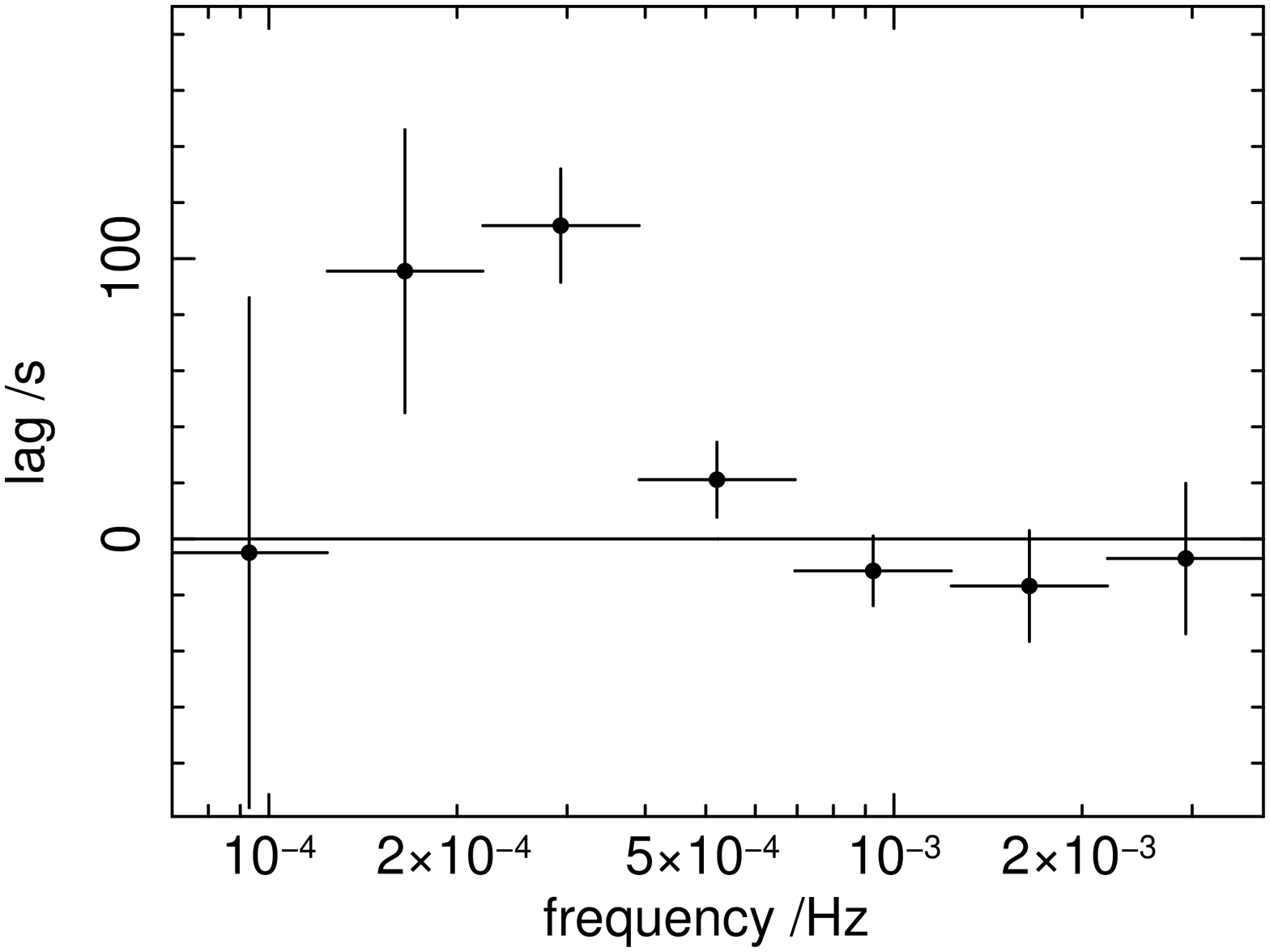}
  \includegraphics[height=.22\textheight,angle=-90]{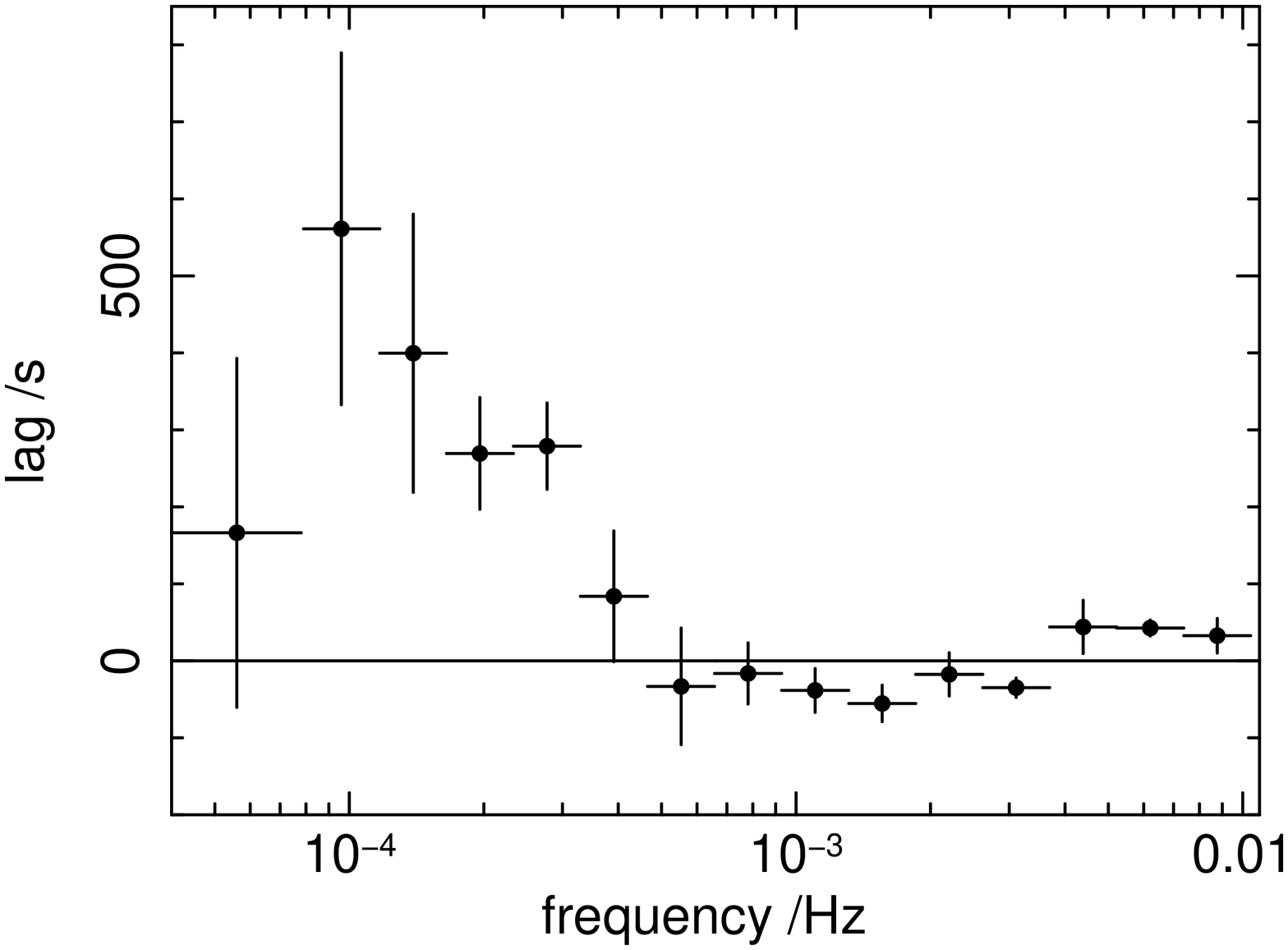}

  \caption{Lag spectra between hard  and soft bands in X-ray observations. Left to right: 
{\it Suzaku} data from  NGC\,4051 \cite{miller10a}; 
{\it XMM-Newton} data from MCG--6-30-15 (Miller et al., in prep;  
also see \citealt{emmanoulopoulos11a}), 
{\it XMM-Newton} data from 1H0707--495 \cite{miller10b}.
Lags are positive when the hard band lags the soft band.}

\end{figure}

Fig.~5 shows some examples of lag spectra having clear transitions from positive lags (hard band lagging
soft) at low frequencies to zero or negative lags at high frequencies.  
The observed transitions are broader 
than  expected for a thin-shell reprocessor, indicating that the reverberating
material has significant depth \cite{miller10a}.  
The observed time delays 
have values of hundreds of seconds at low variation frequencies but fall to values
close to zero at high frequencies.  A typical transition frequency is $\nu \sim 5 \times 10^{-4}$Hz, 
indicating  the reprocessor to be a few light hours from the central source, equivalent to a fews tens
to a few hundreds of gravitational radii, depending on the black hole mass.

The interpretation of large-scale X-ray reverberation is supported by a substantial body of 
spectral evidence \cite{turner09a}. 
Zones of partially-covering, absorbing circumnuclear material shape the X-ray 
timing and spectral properties of AGN. 
 Both spectral and timing analyses indicate that 
the global covering
factor of the reprocessor  must be high ($\sim 50$\%) in order to
achieve a sufficient amount of scattered light to explain the observations. 
Further work is needed to create models that reproduce both
the timing behaviour and the observed high resolution spectra. 


\begin{theacknowledgments}
 We acknowledge NASA grants NNX08AJ41G (TJT) and NNX10AL83H (MT).  

\end{theacknowledgments}



\bibliographystyle{aipproc}   

\bibliography{xray}

\begin{thebibliography}{23}
\expandafter\ifx\csname natexlab\endcsname\relax\def\natexlab#1{#1}\fi
\providecommand{\enquote}[1]{``#1''}
\expandafter\ifx\csname url\endcsname\relax
  \def\url#1{\texttt{#1}}\fi
\expandafter\ifx\csname urlprefix\endcsname\relax\def\urlprefix{URL }\fi
\providecommand{\eprint}[2][]{\url{#2}}

\bibitem[{Gebhardt} et~al.(2000)]{gebhardt00a}
K.~{Gebhardt}, R.~{Bender}, G.~{Bower}, A.~{Dressler}, S.~M. {Faber}, A.~V.
  {Filippenko}, R.~{Green}, C.~{Grillmair}, L.~C. {Ho}, J.~{Kormendy}, T.~R.
  {Lauer}, J.~{Magorrian}, J.~{Pinkney}, D.~{Richstone}, and S.~{Tremaine},
  \emph{\apjl} \textbf{539}, L13--L16 (2000), \eprint{arXiv:astro-ph/0006289}.

\bibitem[{King}(2010)]{king10a}
A.~R. {King}, \emph{\mnras} \textbf{402}, 1516--1522 (2010),
  \eprint{0911.1639}.

\bibitem[{Sim} et~al.(2010)]{sim10a}
S.~A. {Sim}, L.~{Miller}, K.~S. {Long}, T.~J. {Turner}, and J.~N. {Reeves},
  \emph{\mnras} \textbf{404}, 1369--1384 (2010), \eprint{1002.0544}.

\bibitem[{Kaspi} et~al.(2002)]{kaspi02a}
S.~{Kaspi}, W.~N. {Brandt}, I.~M. {George}, H.~{Netzer}, D.~M. {Crenshaw},
  J.~R. {Gabel}, F.~W. {Hamann}, M.~E. {Kaiser}, A.~{Koratkar}, S.~B.
  {Kraemer}, G.~A. {Kriss}, S.~{Mathur}, R.~F. {Mushotzky}, K.~{Nandra}, B.~M.
  {Peterson}, J.~C. {Shields}, T.~J. {Turner}, and W.~{Zheng}, \emph{\apj}
  \textbf{574}, 643--662 (2002), \eprint{arXiv:astro-ph/0203263}.

\bibitem[{Blustin} et~al.(2005)]{blustin05a}
A.~J. {Blustin}, M.~J. {Page}, S.~V. {Fuerst}, G.~{Branduardi-Raymont}, and
  C.~E. {Ashton}, \emph{\aap} \textbf{431}, 111--125 (2005),
  \eprint{arXiv:astro-ph/0411297}.

\bibitem[{Miller} et~al.(2010{\natexlab{a}})]{miller10a}
L.~{Miller}, T.~J. {Turner}, J.~N. {Reeves}, A.~{Lobban}, S.~B. {Kraemer}, and
  D.~M. {Crenshaw}, \emph{\mnras} \textbf{403}, 196--210 (2010{\natexlab{a}}),
  \eprint{0912.0456}.

\bibitem[{Reeves} et~al.(2004)]{reeves04a}
J.~N. {Reeves}, K.~{Nandra}, I.~M. {George}, K.~A. {Pounds}, T.~J. {Turner},
  and T.~{Yaqoob}, \emph{\apj} \textbf{602}, 648--658 (2004),
  \eprint{arXiv:astro-ph/0310820}.

\bibitem[{Turner} et~al.(2008)]{turner08a}
T.~J. {Turner}, J.~N. {Reeves}, S.~B. {Kraemer}, and L.~{Miller}, \emph{\aap}
  \textbf{483}, 161--169 (2008), \eprint{0803.0080}.

\bibitem[{Young} et~al.(2005)]{young05a}
A.~J. {Young}, J.~C. {Lee}, A.~C. {Fabian}, C.~S. {Reynolds}, R.~R. {Gibson},
  and C.~R. {Canizares}, \emph{\apj} \textbf{631}, 733--740 (2005),
  \eprint{arXiv:astro-ph/0506082}.

\bibitem[{Miller} et~al.(2008)]{miller08a}
L.~{Miller}, T.~J. {Turner}, and J.~N. {Reeves}, \emph{\aap} \textbf{483},
  437--452 (2008), \eprint{0803.2680}.

\bibitem[{Risaliti} et~al.(2005)]{risaliti05a}
G.~{Risaliti}, S.~{Bianchi}, G.~{Matt}, A.~{Baldi}, M.~{Elvis}, G.~{Fabbiano},
  and A.~{Zezas}, \emph{\apjl} \textbf{630}, L129--L132 (2005),
  \eprint{arXiv:astro-ph/0508608}.

\bibitem[{Miller} et~al.(2007)]{miller07a}
L.~{Miller}, T.~J. {Turner}, J.~N. {Reeves}, I.~M. {George}, S.~B. {Kraemer},
  and B.~{Wingert}, \emph{\aap} \textbf{463}, 131--143 (2007),
  \eprint{arXiv:astro-ph/0611673}.

\bibitem[{Tombesi} et~al.(2010)]{tombesi10a}
F.~{Tombesi}, M.~{Cappi}, J.~N. {Reeves}, G.~G.~C. {Palumbo}, T.~{Yaqoob},
  V.~{Braito}, and M.~{Dadina}, \emph{\aap} \textbf{521}, A57+ (2010),
  \eprint{1006.2858}.

\bibitem[{Pounds} and {Reeves}(2009)]{pounds09a}
K.~A. {Pounds}, and J.~N. {Reeves}, \emph{\mnras} \textbf{397}, 249--257
  (2009), \eprint{0811.3108}.

\bibitem[{Turner} et~al.(2011)]{turner11a}
T.~J. {Turner}, L.~{Miller}, S.~B. {Kraemer}, and J.~N. {Reeves}, \emph{\apj}
  \textbf{733}, 48--+ (2011), \eprint{1103.3709}.

\bibitem[{Turner} et~al.(2009)]{turner09b}
T.~J. {Turner}, L.~{Miller}, S.~B. {Kraemer}, J.~N. {Reeves}, and K.~A.
  {Pounds}, \emph{\apj} \textbf{698}, 99--105 (2009), \eprint{0903.4347}.

\bibitem[{Reeves} et~al.(2009)]{reeves09a}
J.~N. {Reeves}, P.~T. {O'Brien}, V.~{Braito}, E.~{Behar}, L.~{Miller}, T.~J.
  {Turner}, A.~C. {Fabian}, S.~{Kaspi}, R.~{Mushotzky}, and M.~{Ward},
  \emph{\apj} \textbf{701}, 493--507 (2009), \eprint{0906.0312}.

\bibitem[{Winter} et~al.(2009)]{winter09a}
L.~M. {Winter}, R.~F. {Mushotzky}, C.~S. {Reynolds}, and J.~{Tueller},
  \emph{\apj} \textbf{690}, 1322--1349 (2009), \eprint{0808.0461}.

\bibitem[{Sim} et~al.(2008)]{sim08a}
S.~A. {Sim}, K.~S. {Long}, L.~{Miller}, and T.~J. {Turner}, \emph{\mnras}
  \textbf{388}, 611--624 (2008), \eprint{0805.2251}.

\bibitem[{Patrick} et~al.(2011)]{patrick11a}
A.~R. {Patrick}, J.~N. {Reeves}, D.~{Porquet}, A.~G. {Markowitz}, A.~P.
  {Lobban}, and Y.~{Terashima}, \emph{\mnras} \textbf{411}, 2353--2370 (2011),
  \eprint{1010.2080}.

\bibitem[{Miller} et~al.(2010{\natexlab{b}})]{miller10b}
L.~{Miller}, T.~J. {Turner}, J.~N. {Reeves}, and V.~{Braito}, \emph{\mnras}
  \textbf{408}, 1928--1935 (2010{\natexlab{b}}), \eprint{1006.5035}.

\bibitem[{Emmanoulopoulos} et~al.(2011)]{emmanoulopoulos11a}
D.~{Emmanoulopoulos}, I.~M. {McHardy}, and I.~E. {Papadakis}, \emph{\mnras}
  \textbf{416}, L94--L98 (2011), \eprint{1106.6067}.

\bibitem[{Turner} and {Miller}(2009)]{turner09a}
T.~J. {Turner}, and L.~{Miller}, \emph{\aapr} \textbf{17}, 47--104 (2009),
  \eprint{0902.0651}.

\end{thebibliography}

\IfFileExists{\jobname.bbl}{}
 {\typeout{}
  \typeout{******************************************}
  \typeout{** Please run "bibtex \jobname" to optain}
  \typeout{** the bibliography and then re-run LaTeX}
  \typeout{** twice to fix the references!}
  \typeout{******************************************}
  \typeout{}
 }

\end{document}